\documentclass[preprint,showpacs,amsmath,amssymb]{revtex4}
\usepackage{epsfig }
\usepackage{amsmath}
\usepackage{amssymb}

\newcommand{\p}{\partial}
\DeclareMathAlphabet{\bi}{OML}{cmm}{b}{it}

\begin{document}
\title{Electron transport in waveguides \\ with spatially modulated strengths of the Rashba \\ and Dresselhaus terms of the spin-orbit interaction}
\author{P.  M. Krstaji\'c, E. Rezasoltani, and P. Vasilopoulos}
\address{Concordia University, Department of Physics, 
7141 Sherbrooke Ouest
Montr\'{e}al, Qu\'{e}bec
H4B 1R6, Canada}

\begin{abstract}
We study electron transport through waveguides (WGs)
in which the strengths of the Rashba ($\alpha$) and Dresselhaus ($\beta$) terms of the spin-orbit interaction (SOI) vary in space. Subband mixing, due to lateral confinement, is taken into account only between the two first subbands. For sufficiently narrow WGs the transmission 
$T$  exhibits a square-like shape as a function of $\alpha$ or $\beta$. Particular attention is paid to the case of equal SOI strengths, $\alpha=\beta$,  for which spin-flip processes are expected to decrease. The transmission exhibits resonances as a function  
of the length of a SOI-free region separating two regions with SOI present, that are most pronounced for $\alpha=\beta$. The sign of $\alpha$ strongly affects the spin-up and spin-down transmissions. The results show  that the main effect of subband mixing is to shift the transmission resonances and to decrease  the transmission from one spin state to another. The effect of possible band offsets between regions that have different SOI strengths and effective masses is also
discussed.

\end{abstract}
\maketitle

\vspace{5mm}
\section{Introduction}\label{intro}
There has been growing interest in the studies of spin-orbit interaction (SOI) in 
low-dimensional semiconductor structures made of III-V materials. The spin degree of freedom, often neglected in transport studies in semiconductors like silicon or germanium,  may be important in other materials depending on the crystal structure, growth condition, and 
band alignment of the whole heterostructure.  SOI, of relativistic origin, is a coupling between the intrinsic angular momentum (spin) and the orbital angular momentum  in an external electric field.  SOI manifests itself in semiconductor structures either
due to the lack of (macroscopic) inversion symmetry of the whole structure, referred to as the Rashba SOI term  \cite{bych}, or due to the lack of inversion symmetry of the  crystal structure, referred to as the Dresselhaus SOI term \cite{wink,dres}.
The Rashba   term can also be viewed as an effective magnetic field in the local frame, perpendicular to both momentum and electric field. Apart from the band alignement, it also depends on any external potential if it lifts  the overall inversion symmetry which means it can be tuned by applying a bias \cite{nit,gate2,gate3}. On the other hand, the Dresselhaus SOI (DSOI) term arises as a consequence of the lack of inversion symmetry of 
the underlying crystal structure. It is commonly present in III-V semiconductors, like GaSb that has the zinc-blende structure, where the difference between cations and anions breaks the degeneracy of the band structure with respect to the spin degree of freedom, and is present in both bulk materials and semiconductor nanostructures. In low-dimensional semiconductor structures the DSOI manifests itself through  terms that are  linear and cubic  in the wave vector $k$; here we consider only the former, which is  dominant for small $k$ and is referred to as the $[001]$ linear Dresselhaus term. There is an additional source of spin splitting present in semiconductor heterostructures due to the reduced symmetry at the interface  \cite{Ivch,Krebs}. This manifestation of spin-orbit coupling is often named interface inversion symmetry or interface Dresselhaus SOI \cite{Gan2}. 

The studies of spin-dependent phenomena in semiconductor structures have been particularly intensified  
after the proposal of a spin-field effect transistor (FET) by Datta and Das \cite{Datta}. This kind of the device would make use of the Rashba SOI only, by controlling the electron spin
during its passage through the transistor. Ever since this proposal,  there have been many  
refinements of the idea, notably the non-ballistic spin field-effect transistorÊ\cite{schl1} which would utilize both the Rashba and Dresselhaus terms of  
equal strength. In this design it is expected that the transistor is robust against spin-indepedent scattering mechanisms. Further, a modification of the Datta-Das device has been proposed  \cite{SFET_DSO}
whose function would be based on solely the DSOI.  
Motivated by this idea a wealth of related studies appeared in similar systems that dealt with
spin-dependent transport, see, e.g, the review paper \cite{zut}.

 In previous work, coauthored by one of us, ballistic transport and spin-transistor behavior was studied, due only to the RSOI, in stubbed \cite{wan1} WGs with constant strength $\alpha$ or  unstubbed WGs with periodically modulatedÊ$\,\alpha$ \cite{wan2}. An encouraging result was a nearly square-wave form of the transmission as a function of some stub parameters \cite{wan1} or the strength \cite{wan2} $\alpha$.  In this work  we extend these studies  
by treating simultaneously both SOI terms, taking into account mixing between the lowest two subbands, and by studying  
{\it longitudinal} transmission resonances that occur when the length 
of a SOI-free region, separating two regions with SOI present, varies. As will be seen, if only one subband is occupied and both SOI terms are present, a phase difference $\phi=\tan^{-1}(-\beta/\alpha)$ arises in the spin eigenfunctions that strongly affects 
the spin-up and spin-down transmissions especially when $\phi$ changes sign. In Sec.~\ref{theory} we present a model  
of a WG 
with two subbands, due to a lateral confinement, having nonzero mixing. We also derive the relevant  dispersion relations and one-electron wave functions. In Sec.~\ref{numres}  we briefly  explain the numerical procedure  and present the  main results. 
 Concluding remarks follow in Sec.~\ref{conc}.

\section{Theoretical model}\label{theory}
\begin{figure}[h]
\begin{center}
\vspace*{-0.5cm}
 \includegraphics[height=3.5cm, width=8cm]{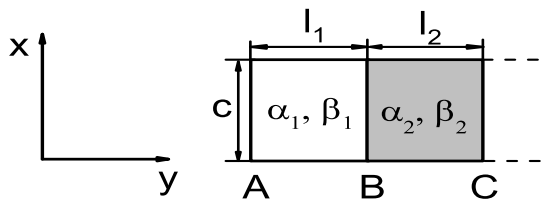} 
\end{center}
\vspace{-1.19cm}
\caption{\label{fig1} 
Schematics of a WG unit, of width $c$, consisting of two segments with lengths $l_1$, $l_2$, and SOI strengths $\alpha_i$, $\beta_i$,  
($i=1,2$). The full unit is  repeated along the $y$ direction.}   
\end{figure}

 One unit of the WG we consider  is shown  
 in Fig.~\ref{fig1}. It is made,   e.g.,  of two layers of In$_x$Ga$_{1-x}$As with different content of In, $x_i$, and   has SOI strengths $\alpha_i$ and $\beta_i$. The WG consists of a finite number of such units periodically repeated in the $y$ direction. Along the $x$ direction a confining potential $V(x)$ is present that gives rise to bound states having energies $E_n$. 
 In principle  a  confinement along $x$ could be created by etching after the usual 2D one along $z$. 
 The two different growth directions   which will be discussed are $[010]$ and $[110]$. 
 \subsection{$[010]$ grown waveguide}
 The one-electron Hamiltonian reads 
\begin{equation}\label{ham}
H=H_0+H_{\alpha}+H_{\beta}\,,
\end{equation}
\noindent where $H_0$ is given by
\begin{equation}\label{ham1}
H_0 = \frac{\hbar^2{\bf k}^{2}}{2m^{\ast}} +V(x)\,.
\end{equation}
\noindent  Here ${\bf k}=(k_x,k_y)$ is the wavevector of the electron  and
$m^\ast$ its effective mass. $H_{\alpha}$ and $H_{\beta}$ are the Rashba and Dresselhaus terms, respectively, given by
\begin{equation}\label{ham23}
H_\alpha =\alpha [\sigma_x k_y 
-\sigma_y k_x],\,  
\,\,H_\beta =\beta [\sigma_x k_x 
-\sigma_y k_y]\,, 
\end{equation} 
where  ${\bf \sigma}=(\sigma _{x},\sigma _{y},\sigma _{z})$
are the Pauli spin matrices, and $\alpha$,  
$\beta$ the strengths of the Rashba and Dresselhaus terms, respectively. 
In writing Eq.~(\ref{ham23}) we assumed that the strength $\beta$ depends mainly on the confinement along the $z$ axis, 
only the linear-in-wave vector Dresselhaus term is important, and neglected any dependence on the lateral
confinement along the $x$ axis. The idea of linearly changing $\beta$ by changing the well width along $z$ has recently been reconfirmed in Ref.~\onlinecite{jak}.

We write the total wave function  
as a linear combination of eigenstates of the unperturbed Hamiltonian
\begin{equation}\label{wavefx}
\Psi(x,y)=\sum_{n, \sigma} A_n^\sigma\phi_n(x)|\sigma\rangle e^{ik_yy}\,,
\end{equation}
with  $n$ labelling the discrete subbands $E_n$ due to  
the confining potential $V(x)$.  
The unperturbed states satisfy $H_0|n, k_y, \sigma\rangle=E^0_n|n, k_y, \sigma\rangle$, with 
$E^0_n=E_n+\lambda k_y^2$ and $ \lambda=\hbar^2/2m^*$.  $\phi_n(x)$ is the
solution of 
\begin{equation}
\big[-\lambda \frac{d^2}{dx^2} +V(x)\big]\phi_n(x)=E_n \phi_n(x)\,,
\end{equation}
with the square-type $V(x)$  assumed  
high enough  so that $\phi_n(x)=0$ at the edges of the WG. 
\begin{figure}[h]
\begin{center}
\hspace*{-0.3cm}
 \includegraphics[height=4.8cm, width=6.8cm]{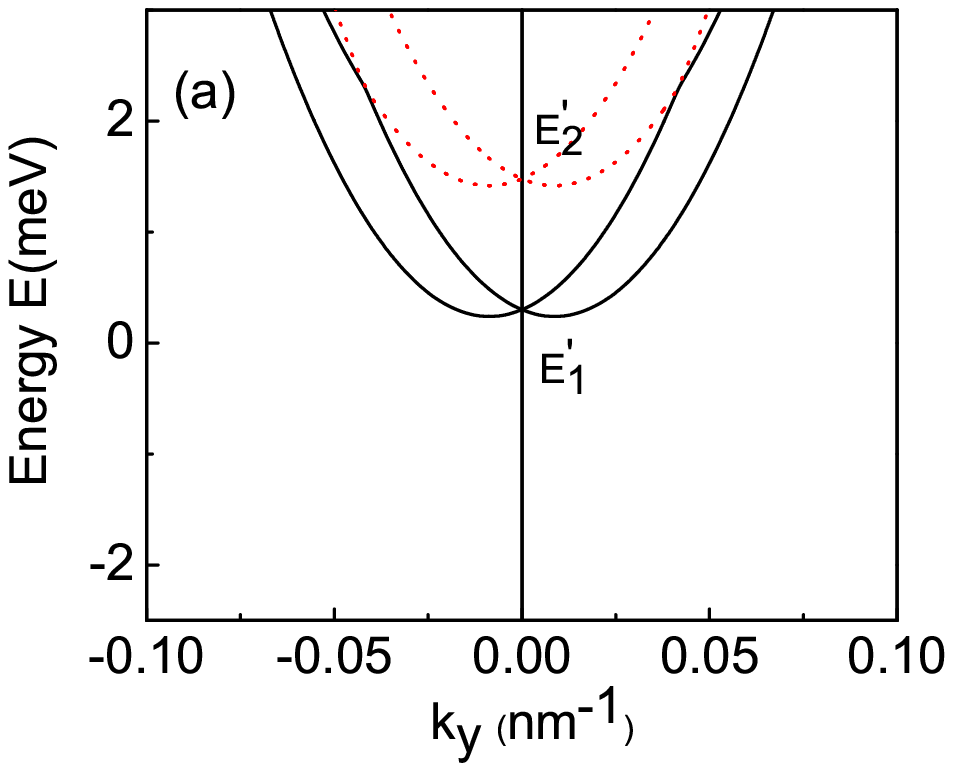} 
\hspace*{-0.3cm}
  \includegraphics[height=5cm, width=6.8cm]{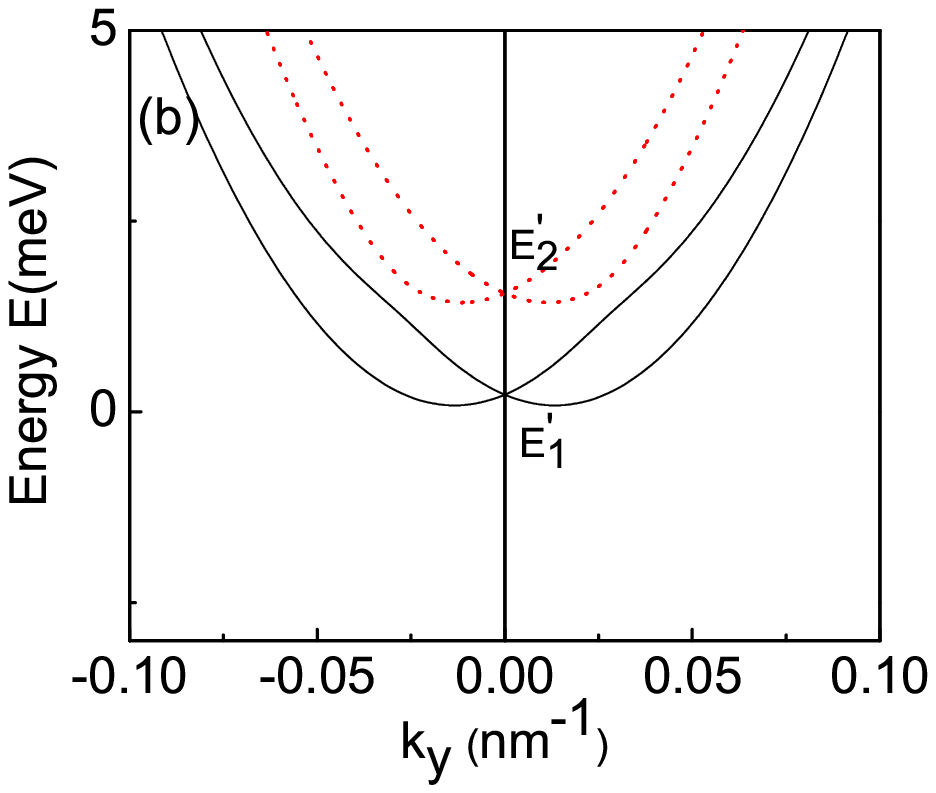} 
\end{center}
\vspace{-1.29cm}
\caption{\label{fig2} 
Dispersion relation $E(k_y)$ for a WG, with equal SOI 
strengths  $\alpha=\beta= \alpha_0=1\times10^{-11}$ eVm in (a) 
and for $\alpha=\alpha_0$, $\beta=2\alpha_0$, in (b), where an anticrossing of the curves 
is visible. The interesections of the curves with the $y$ axis, at $E_{1,2}^{\prime}$, 
are close  to the subband energies $E_{1,2}$ as explained in the text.}
\end{figure}

Using the full Hamiltonian given by Eqs.~(\ref{ham})-(\ref{ham23}) and setting $\bar{E}_n=E_n^0-E$ and 
$\gamma_\pm =\alpha\pm i\beta$, Eq.~(\ref{wavefx}) leads to the  
secular equation  
\begin{equation}
\hspace*{-0.2cm}
\left(
\begin{array}{cc}
\bar{E}_n  
\, \, \,\,\quad\gamma_+k_y\\
\ \\
\gamma_-k_y \, \, \,  \quad\bar{E}_n  
\end{array}
\right)
\hspace*{-0.1cm}
\left(
\begin{array}{cc}
A_n^{+}\\
\ \\ A_n^{-}
\end{array}
\right)
+\sum_{m}J_{nm} 
\left(
\begin{array}{cc}
0\, \, \quad\gamma_- \\
\ \\
-\gamma_+ \, \,\, 0\end{array}\right)
\hspace*{-0.13cm}
\left(
\begin{array}{cc}
A_m^{+}\\
\ \\
A_m^{-}
\end{array}
\right)=0\,.
\label{subbeq}
\end{equation}
The factor  $J_{nm}$ embodies the subband mixing due to confinement 
and is nonzero for $n\neq m$, 
\begin{equation}
J_{nm}=\int \phi_n(x)\phi_m^{\prime}(x)dx\,.
\end{equation}
 
To assess the magnitude of subband mixing  we assume that the confining potential $V(x)$ is that of a quantum well with infinitely high walls at the edges of the WG whose  width is $c$. Then  $\phi_{n}(x)=(2/c)^{1/2}\sin(n\pi x/c)$ 
and considering mixing between the lowest two subbands the only non-vanishing matrix elements  are $J_{12}=-J_{21}=\delta$, with $\delta=-8/(3c) $ \cite{wan1}. Under these 
assumptions the eigenvalue problem resulting from Eq.~(\ref{subbeq}) takes the  form
\begin{equation}\label{Eig}
\left(
\begin{array}{cccc}
E_1^{0} & \gamma_{+}k_y & 0 &\gamma_{-}\delta \\
\gamma_{-}k_y & E_1^{0} & -\gamma_{+}\delta & 0 \\
0 & -\gamma_{-}\delta & E_2^{0} & \gamma_{+}k_y \\
\gamma_{+}\delta & 0 & \gamma_{-}k_y & E_2^{0}
\end{array}
\right)
\left(
\begin{array}{c}
A_1^{+} \\
A_1^{-} \\
A_2^{+} \\
A_2 ^{-}
\end{array}
\right)=E\left(
\begin{array}{c}
A_1^{+} \\
A_1^{-} \\
A_2^{+} \\
A_2^{-} 
\end{array}
\right)\,,
\end{equation}
\noindent where $E_1^{0}=E_1+\lambda k_y^2,\,\,E_2^{0}=E_2+\lambda k_y^2$.  
The  eigenenvalues, readily found from Eq.~(\ref{Eig}),  are 
\begin{subequations}
\begin{equation}
\varepsilon_1^{+}= 
\big[E_1^{0}+E_2^{0}-[G-F\,k_y]^{1/2}\big]\big/2\,,
\end{equation}
\begin{equation}
\varepsilon_1^{-}= 
\big[E_1^{0}+E_2^{0}-[G+F\,k_y]^{1/2}\big] \big/2\,,
\end{equation}
\begin{equation}
\varepsilon_2^{+}= 
\big[E_1^{0}+E_2^{0}+[G+F\,k_y]^{1/2}\big] \big/2\,,
\end{equation}
\begin{equation}
\varepsilon_2^{-}= 
\big[E_1^{0}+E_2^{0}+[G-F\,k_y]^{1/2}\big] \big/2,
\end{equation}
\end{subequations}

\noindent with $F=4[16\alpha^2\beta^2\delta^2+\gamma^2\Delta E_{12}^2]^{1/2}$, $G=4\gamma^2(\delta^2+k_y^2)+\Delta E_{12}^2$ ($\gamma^2=\alpha^2+\beta^2$), and $\Delta E_{12}=E_2-E_1$. The energy dispersions are given in the left panel of Fig.~\ref{fig2} for 
$\alpha=\beta=\alpha_0$ and 
in the right one for $\alpha=0.5\beta=2\alpha_0$. 

Note that the energy dispersion curves do not start from 
$E_1$, $E_2$ at ${\bf k}=0$ but rather from
\begin{subequations}
\begin{equation}
E_1^{\prime}= 
\big[E_s 
-[4\gamma^2\delta^2+\Delta E_{12}^2]^{1/2}\big]\big/2\,,
\end{equation}
\begin{equation}
E_2^{\prime}= 
\big[E_s 
+[4\gamma^2\delta^2+\Delta E_{12}^2]^{1/2}\big]\big/2\,,
\end{equation}
\end{subequations}

\noindent where $E_s=E_1+E_2$, as a result of the subband mixing. 

Analytical expressions for the wavevector $k_y(E)$ as a function of the energy  are complicated for the general case $\alpha\neq\beta$. 
Particular attention will be paid to the case $\alpha=\beta$ 
in which a suppression of spin-flip 
processes is expected \cite{schl1,Golub}. The relevant expressions are 
\begin{subequations}
\begin{equation}\label{disp1}
k^{\pm}_{y1}=
\Big[\mp\alpha+[\alpha^2-\lambda(E_s 
-D-2E)]^{1/2}\Big]/\sqrt{2}\lambda,
\end{equation}
\begin{equation}\label{disp2}
k^{\pm}_{y2}=
\Big[\alpha\mp[\alpha^2-\lambda(E_s 
+D-2E)]^{1/2}\Big]/\sqrt{2}\lambda,
\end{equation}
\end{subequations}

\noindent where  
$D=(8\alpha^2\delta^2+\Delta E_{12}^2)^{1/2}$. 
From Eqs. (\ref{disp1})-(\ref{disp2})  one can derive 
the critical energies 
\begin{eqnarray} 
E_{cr1} &=&  (E_s -D)/2- \alpha^2/2\lambda,\\
E_{cr2} &= &  (E_s +D)/2- \alpha^2/2\lambda\,,
\end{eqnarray} 

\noindent that determine the nature of the wave 
vectors $k_{yi}$ in the following manner:

\begin{widetext}
\noindent $\quad E\leq E_{cr1}:\quad\quad$ for fixed $i$ ($i=1,2$) all solutions $k_{yi}$ are complex  and conjugate in pairs; \\
$E_{cr1}<E< E_{cr2}:\,$ the solutions $k_{y1}$ are real  and the $k_{y2}$ ones complex conjugate;\\
$\quad\quad\quad E\geq E_{cr2}:\quad\quad\quad$ all solutions $k_{yi}$ are real.
\end{widetext}
For vanishing Dresselhaus strength $\beta\rightarrow0$, the eigenvalue problem Eq.~(\ref{Eig}) simplifies significantly and the eigenvectors 
acquire \cite{Wangsubbmix} the  simple  analytical form
\begin{subequations}
\begin{equation}
\Psi_1^{+}=\frac{1}{C}\left
(
\begin{array}{c} 
1 \\
1 \\
r_B \\
-r_B
\end{array}\right), \quad 
\Psi_1^{-}=\frac{1}{D}\left(
\begin{array}{c} 
-1 \\
1 \\
r_A \\
r_A
\end{array}\right), \quad 
\end{equation}
\begin{equation}
\Psi_2^{+}=\frac{1}{D}\left(
\begin{array}{c} 
r_A \\
-r_A \\
1 \\
1
\end{array}\right), \quad 
\Psi_2^{-}=\frac{1}{C}\left(
\begin{array}{c} 
r_B \\
r_B \\
-1 \\
1
\end{array}\right), \quad 
\end{equation}
\end{subequations}
\noindent where $r_A = 2\alpha\delta/A$, $r_B = 2\alpha\delta/B$, $A = \delta E_{12}+2\alpha k_y+\Delta E_{+}$, and $B =\delta E_{12}+2\alpha k_y+\Delta E_{-}$. Further, 
$\Delta E_{\pm}=[(\Delta E_{12}\pm2\alpha k_y)
^{2}+4\alpha^2\delta^2]^{1/2}$, $D=(2+2r_A^2)^{1/2}$, and $C=(2+2r_B^2)^{1/2}$.

If one goes further and neglects subband mixing, by taking the limit $\delta\rightarrow0$, and if only the first subband is occupied, the original $4\times 4$ eigenvalue problem, Eq.~(\ref{Eig}),  
reduces essentially to a $2\times2$ problem. Then the energy spectrum is given by  
\begin{equation}\label{energydisp}
\varepsilon^{\pm}=E_1+\lambda k_y^2\pm(\alpha^2+\beta^2)^{1/2}k_y
\end{equation} 
and the spinors 
acquire the simple form
\begin{equation}\label{eigvec2}
 \psi_e=\frac{1}{\sqrt{2}}\left
 (\begin{array}{c}
 1 \\
 \pm e^{i\phi}
 \end{array}  \right
 ), \,\tan\phi=-\beta/\alpha. 
\end{equation}
This form of the spinors  
is important for the analysis of the transport problem through WGs. 
More precisely, one easily 
sees that the effect of the  
presence of the DSOI term is not just a simple increase of the overall SOI coupling,
that is, $[\alpha^2+\beta^2]^{1/2}$ in place of $\alpha$; one also has the change in the phase of  the spinor component, that may significantly alter the transmission from one spin state 
to another. For illustrative purposes, we  derive an analytical expression for the total transmission  through a simple WG segment, with  equal SOI strengths ($\alpha=\beta$) and length $\ell_2$, sandwiched between two SOI-free segments. The result is
 \begin{equation}\label{single}
T_x=\frac{1+\cos^2\epsilon}{2(1+r\sin^2\Delta_2\ell_2)},
\end{equation} 
\noindent where $\epsilon=\alpha\ell_2/\sqrt{2}\lambda $, $r=(\Delta_1^2-\Delta_2^2)^2/4\Delta_1^2\Delta_2^2$,  $\Delta_1 =[4(E-E_1)\lambda]^{1/2}/2\lambda$, and $\Delta_2 =[2\alpha^2+4(E-E_1)\lambda]^{1/2}/2\lambda$. 
Once again, the effect of having both SOI terms present is not limited to the replacement $\alpha\rightarrow\sqrt{2}\alpha$;  the transmission amplitude is also modulated through the factor $\epsilon$ in Eq.~(\ref{single}), if one compares with the simplest case ($\beta=0, \alpha\neq 0$) \cite{wan2}. From now on  we will evaluate the  transmission  of spin states 
with $z$ being the quantization axis.  

\subsection{ Waveguide grown along the $[110]$ direction }\label{SecthB}
\begin{figure}[h]
\begin{center}
\hspace*{-0.3cm}
 \includegraphics[height=5cm, width=6.8cm]{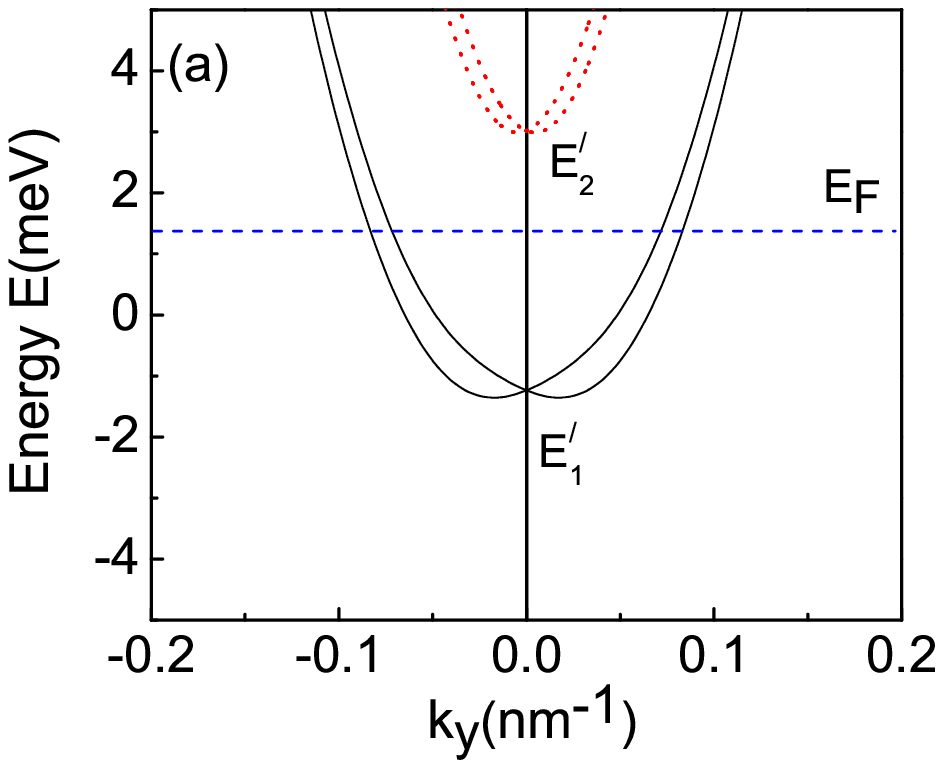} 
\hspace*{-0.3cm}
  \includegraphics[height=5cm, width=6.8cm]{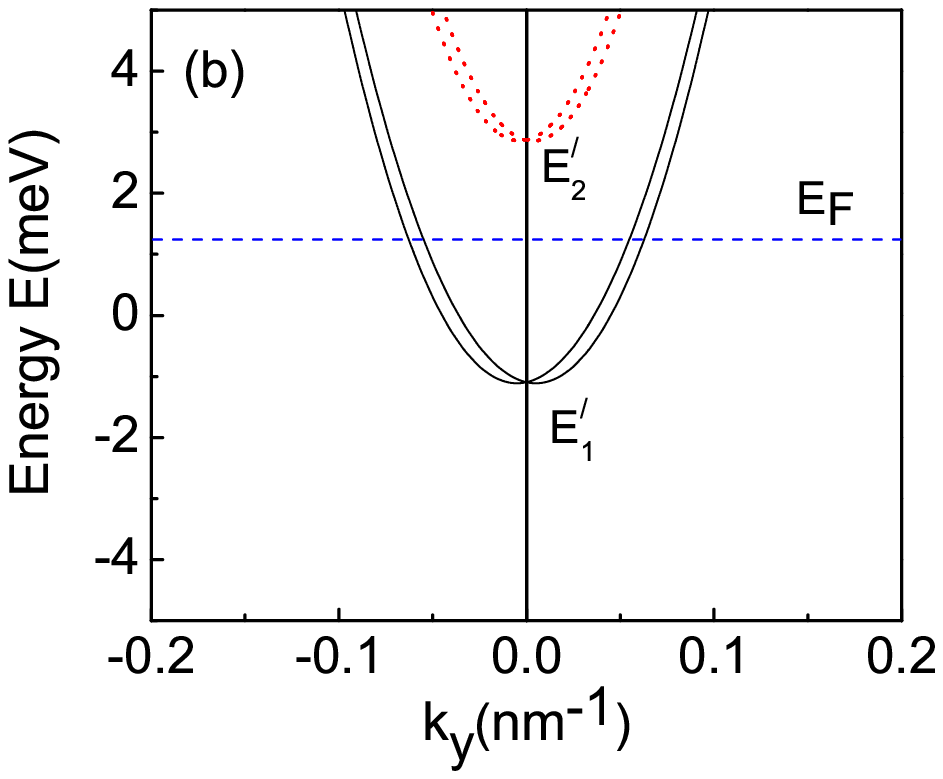} 
\end{center}
\vspace{-1.19cm}
\caption{\label{fig3110} 
Dispersion relation $E(k_y)$ for a WG grown along the [110] direction with equal SOI 
strengths in (a),  $\alpha=\beta= \alpha_0=1\times10^{-11}$ eVm,  
and unequal ones in (b), $\alpha=\alpha_0$, $\beta=2\alpha_0$.}
\end{figure}

 Apart from the usual growth direction along the crystallographic axis $<001>$, the growth along the  Ê$[110]$ direction is also important to investigate. 
 The Dresselhaus  term $H_{\beta}$ has a somewhat simpler form 
 \cite{Dres110}
\begin{equation}\label{hamD110}
H_\beta^{([110])} =-2\beta\sigma_zk_x\,, 
\end{equation} 
\noindent while the Rashba term retains the same form since it is related to the structural 
(macroscopic) inversion asymmetry \cite{anothWang}. Employing a procedure similar to that in Sec. III A, 
we arrive 
at the 
eigenvalue problem 
\begin{equation}\label{Eig}
\left(
\begin{array}{cccc}
E_1^{0} & \alpha k_y & 2i\beta\delta &\alpha\delta \\
\alpha k_y & E_1^{0} & -\alpha\delta & -2i\beta\delta \\
-2i\beta\delta & -\alpha\delta & E_2^{0} & \alpha k_y \\
\alpha\delta & 2i\beta\delta & \alpha k_y & E_2^{0}
\end{array}
\right)
\left(
\begin{array}{c}
A_1^{+} \\
A_1^{-} \\
A_2^{+} \\
A_2 ^{-}
\end{array}
\right)=E\left(
\begin{array}{c}
A_1^{+} \\
A_1^{-} \\
A_2^{+} \\
A_2^{-} 
\end{array}
\right)\,,
\end{equation}
\noindent where the notation is the same as in Eq. (8).

The  eigenenvalues, readily found from Eq.~(\ref{Eig}), are 
given by Eq. (9) with $F$ and $G$ replaced, respectively, by $F_1=4\alpha\Delta E_{12}$ and  $G_1=4\alpha^2(\delta^2+k_y^2)+\Delta E_{12}^2 +16\beta^2\delta^2$. 
The energy dispersions are given on the left panel of Fig.~\ref{fig3110} for 
$\alpha=\beta=\alpha_0$ and on the right one for $\alpha=0.5\beta=2\alpha_0$. The dispersion curves are qualitatively different than those pertaining to 
 the $[010]$ direction due to the different form of the Dresselhaus term $H_{\beta}$. 
Further, the difference between 
equal and unequal SOI strengths is not as drastic as for WGs grown along the $[010]$ direction.  

\begin{figure}[h]
\begin{center}
\hspace*{-0.5cm}
   \includegraphics[height=6cm, width=10cm]{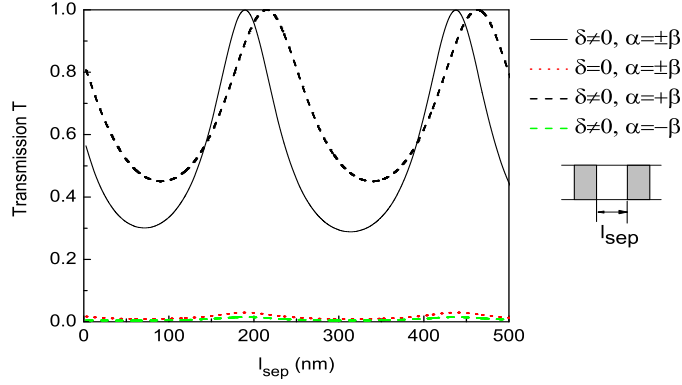}
\end{center}
\vspace*{-1.19cm}
\caption{\label{fig3} Transmission through two  
segments, with $\alpha=\beta=2\alpha_0$, versus the length $\ell_{sep}$ of the SOI-free region that separates them, for fixed $\ell_1=\ell_2=95$nm.   
The solid curve corresponds to the total transmission while
the dotted one to $T^{+-}$ for  
$\alpha=\beta$ in both segments and the dash-dotted one to $T^{+-}$ for $\alpha=-\beta< 0$ in the second segment. 
For comparison the total $T$ is shown for zero subband mixing by  the dashed curve. Here $\alpha_0=1\times10^{-11}$eVm.}
\end{figure}

\section{Numerical procedure and results}\label{numres}
As shown in Fig.~\ref{fig1}, the WG consists of two segments,  with different values of the Rashba and Dresselhaus couplings $\alpha$ and $\beta$, 
periodically repeated  along the $y$ axis. In each of the segments the wave vector $k_y$ is constant so that the wave function $\varphi_i(x,y)$ in the $i$-th segment 
is a superposition of plane wave-like states 
consisting 
of the eigenvectors $d_{j}$ ($j=1,..,4$) of Eq.~(\ref{Eig}), in both directions along the $y$ axis
\begin{equation}\label{wavefun2}
\vspace*{-0.4cm}\varphi_i(x,y)=\sum_{j}\left[c_{i}^{(j)}\cdot \psi_{j} e^{ik_{yj}y_s}+\bar{c}_{i}^{(j)}\cdot \bar{\psi}_{j}e^{-ik_{yj'}y_s}\right](2/c)^{1/2}\sin(n\pi x/c),
\end{equation}
\noindent where $y_s=y-y_{0i}$. To find the complete solution, we  
first match the wave function at the interfaces between the $i$ and $i +1$ segments. Due to the presence of the off-diagonal elements in the Hamiltonian the continuity 
of the derivative of the wave function may not hold. A more general procedure is to require that the flux through materials with different SOI strengths  or/and effective 
masses be conserved \cite{mol}. The velocity operator is given by
 \begin{equation}
v_y={\p H\over \p p_y}={1\over \hbar}
\left[
\begin{array}{cccc}
-2i\lambda\p/\p y & \gamma_+ & 0 & 0 \\
\gamma_- & -2i\lambda\p/\p y & 0 & 0 \\
0 & 0 & -2i\lambda\p/\p y & \gamma_+ \\
0 & 0 & \gamma_- & -2i\lambda\p/\p y
\end{array}\right]\,.
\end{equation}

\noindent The continuity of the wave function at the interface $y=y_{i,i+1}$
gives  $\varphi_{i+1}(x, y_{i,i+1})=\varphi_{i}(x, y_{i,i+1})$ and that of the flux
$\hat{v}_y\varphi_{i+1}(x, y)|_{y_{i,i+1}}=\hat{v}_y\varphi_{i}(x, y)|_{y_{i,i+1}}$.  
The unknown coefficients $c_{i}^{(j)}$ from Eq.~(\ref{wavefun2}) from one segment to  another can be related through the transfer-matrix formalism by introducing 
the propagation matrix $P_i$ and the boundary matching matrix $Q_i$ in each segment $i$. The transfer matrix \cite{Xu} for the $i$-th segment is the matrix  product 
\begin{equation}\label{TMM}
M(i,i+1) = P_i^{-1}Q_i^{-1}Q_{i+1}\,.
\end{equation}
\begin{figure}[h]
\begin{center}
\setlength{\unitlength}{1cm}%
\hspace*{-0.4cm}  \includegraphics[height=6.4cm, width=6.6cm]{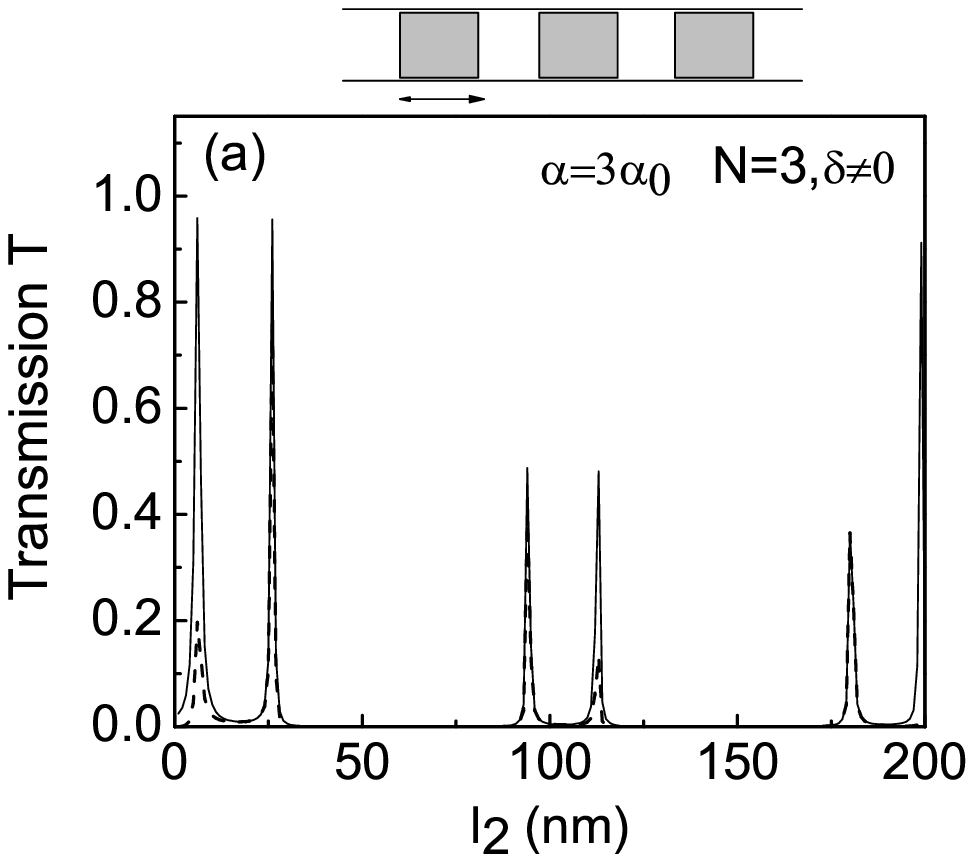}
\hspace*{-0.5cm}
  \includegraphics[height=6.4cm, width=6.6cm]{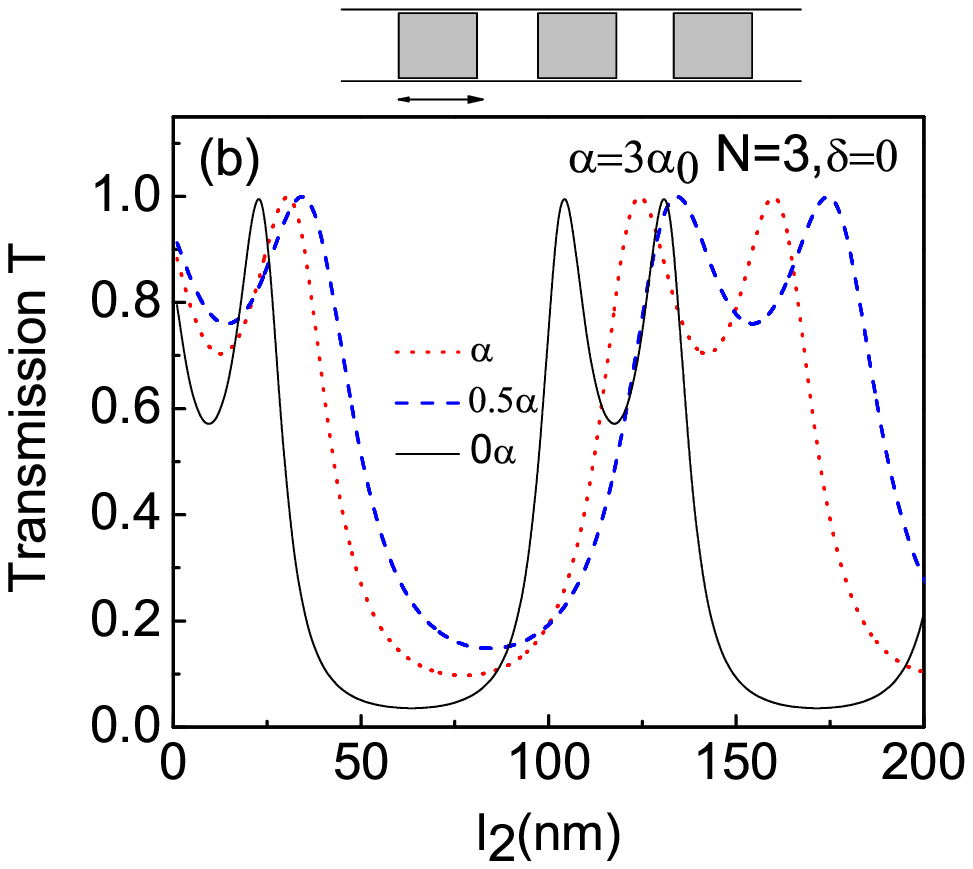}
  \end{center}
\vspace*{-1.2cm}
\caption{\label{fig4} (a) Transmission through three units versus length $\ell_2$, for fixed $\ell_1=95$nm and $\beta=\alpha=3\alpha_0$. 
 The solid curve is the total transmission and the  dashed curve the $T^{+-}$ one.
(b) Same as in (a) but with zero subband mixing and  
three values of $\beta$: $\beta=\alpha=3\alpha_0$ (dotted red curve), $\beta=0.5\alpha$ (dashed blue  curve), and $\beta=0$ (solid black curve). } 
\end{figure}
In all numerical calculations we assumed that the incident electrons are (spin) unpolarized  
and we investigate only the transmission of one spin state, for instance, the spin-up one.   
We take the $z$ axis as the quantization axis.  
We measure the SOI strengths in units of $\alpha_0=1\times10^{-11}$ eVm (Ref.~\cite{gate3}) and we first 
consider an energy $E=0.13$  meV+$E_1$  below the second subband. 
\begin{figure}
\begin{center}
 \includegraphics[height=5.4cm, width=7.8cm]{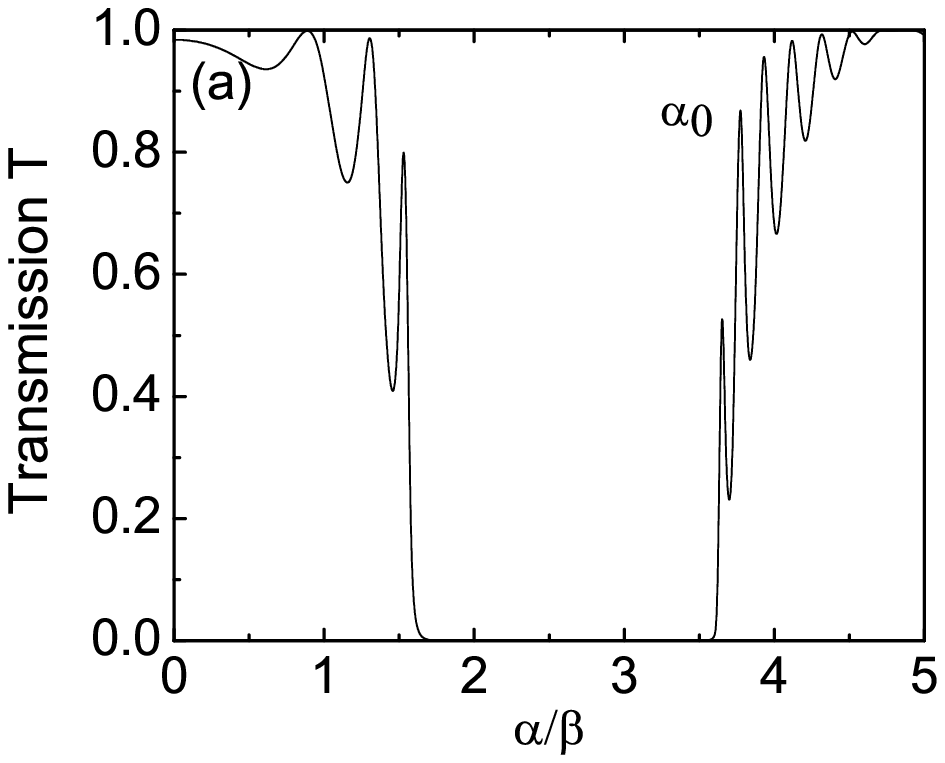}
  \includegraphics[height=5.4cm, width=7.8cm]{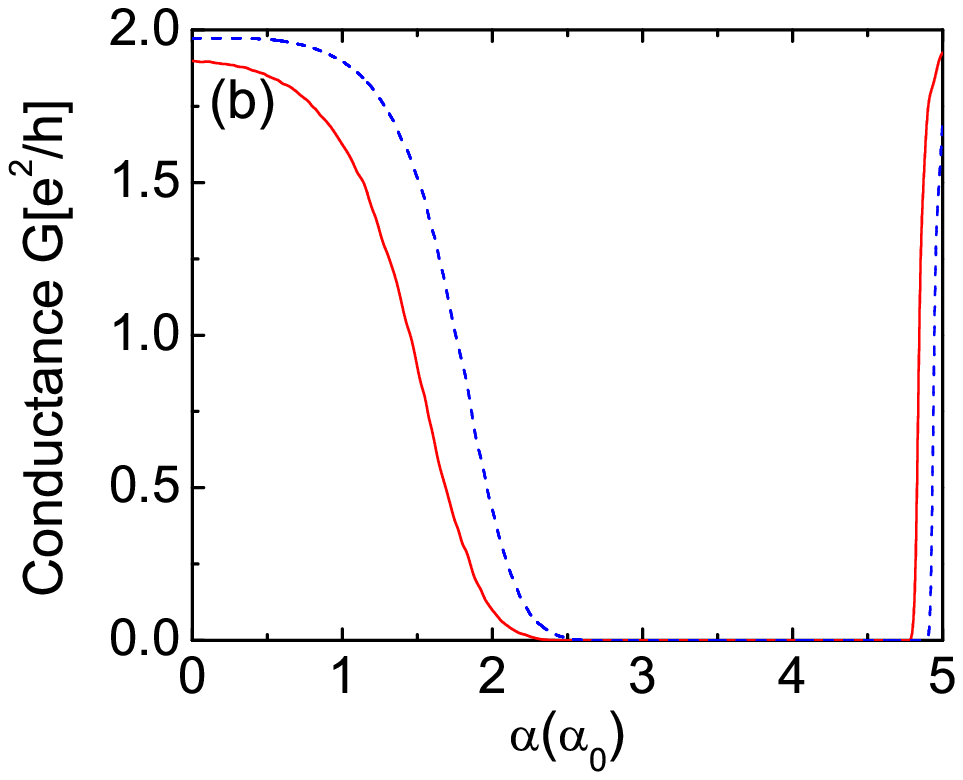}
   \includegraphics[height=5.4cm, width=7.8cm]{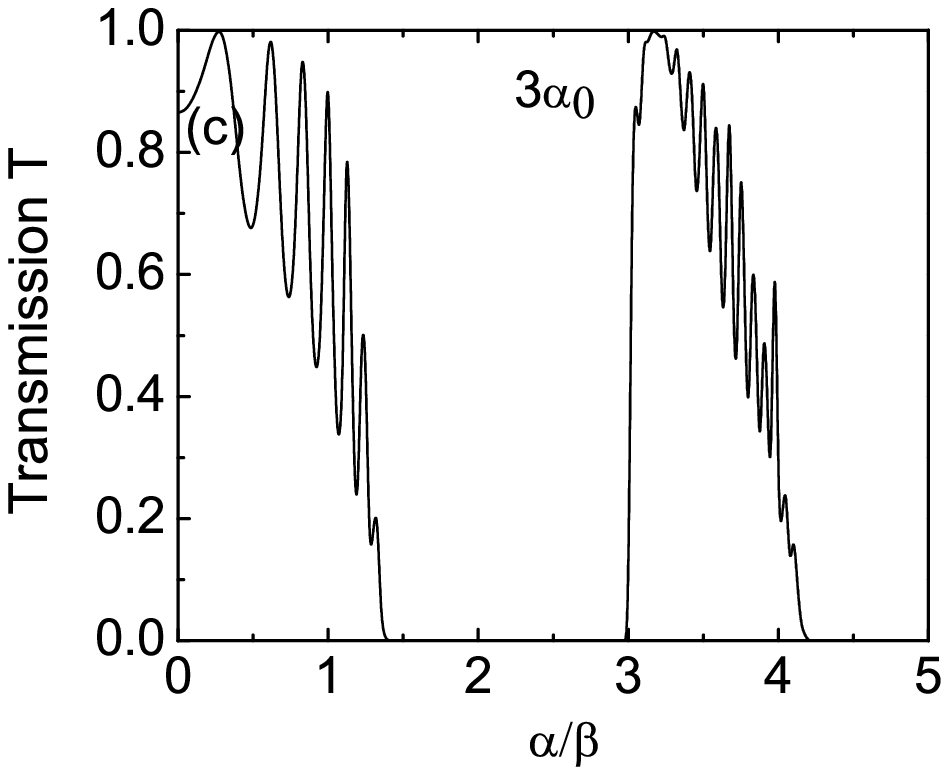}
   \includegraphics[height=5.4cm, width=7.8cm]{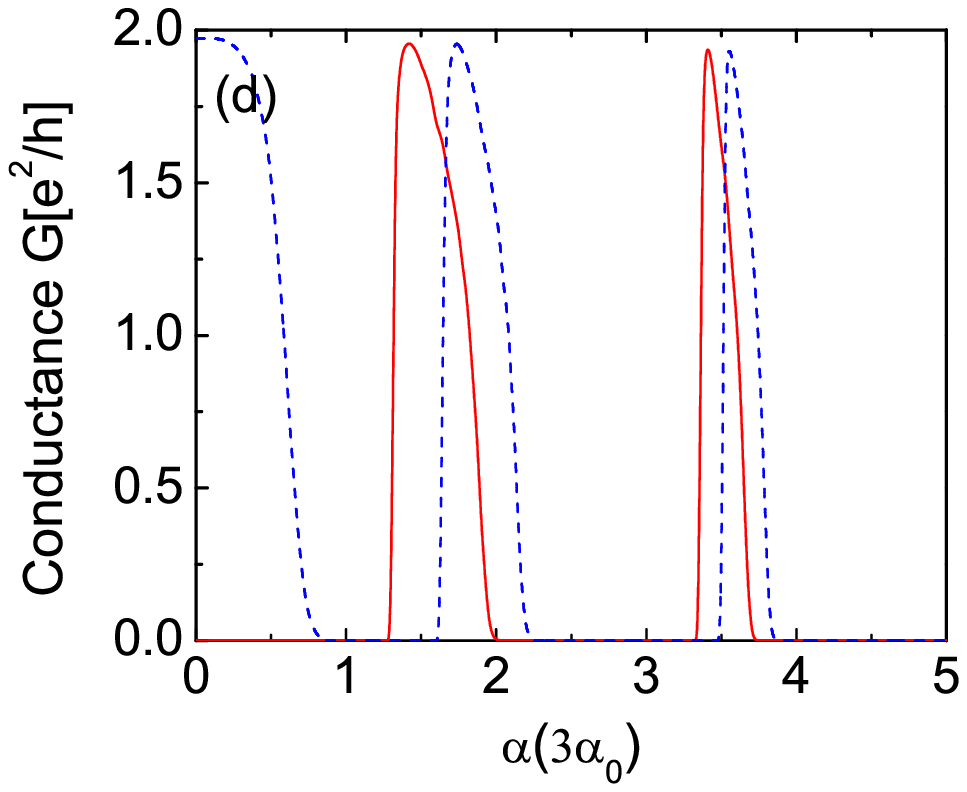}
\end{center}
\vspace*{-0.99cm}
\caption{\label{fig5} (a), (c): Transmission through $20$ units versus the ratio $\alpha/\beta$ with the SOI strengths given, respectively, in units 
of $\alpha_0$ and $3\alpha_0$. 
(b), (d): Conductance $G$, at finite temperature $T=0.2$K, versus 
$\alpha$ with $\beta$ kept constant and equal, respectively, to $\alpha_0$  
 and $3\alpha_0$ (solid red curves),  and $\beta=0$ (dashed blue curves). 
In all cases the unit lengths are $\ell_1=\ell_2=95$ nm, the energy $E=0.13$ meV, and the SOI is absent in the first segment of the unit.} 
\end{figure}

First, we investigate the electron transmission through two segments having equal SOI strengths $\alpha=\beta$ 
and being separated by a SOI-free region of length $\ell_{sep}$. The results are shown in Fig.~\ref{fig3} for several cases: (1) $\alpha$ and $\beta$ 
have the same sign in both segments; the total transmission is shown by the solid curve and the transmission from the spin-up to the spin-down state by the dotted curve;
(2) the same as in  case (1) but with negligible subband mixing ($\delta\to 0$,  solid dashed curve); (3) $\alpha=-\beta$ in the second segment and $T^{+-}$  shown by the dash-dotted curve. We note that the total transmission $T$ is the same in cases (1) and (3), only the spin-up and spin-down transmissions are different in these cases. Here the value of $\alpha$ is taken to be $\alpha=2\alpha_0$,
 close to the experimental value given in Ref.~\cite{nit}. It can be inferred that 
changing the sign of $\alpha$ suppresses the transmission to the opposite spin state, while the subband mixing shifts the resonance maxima and has a minimal effect on the shape of the curve. 
\begin{figure}[b]
\begin{center}
 \includegraphics[height=6.4cm, width=8cm]{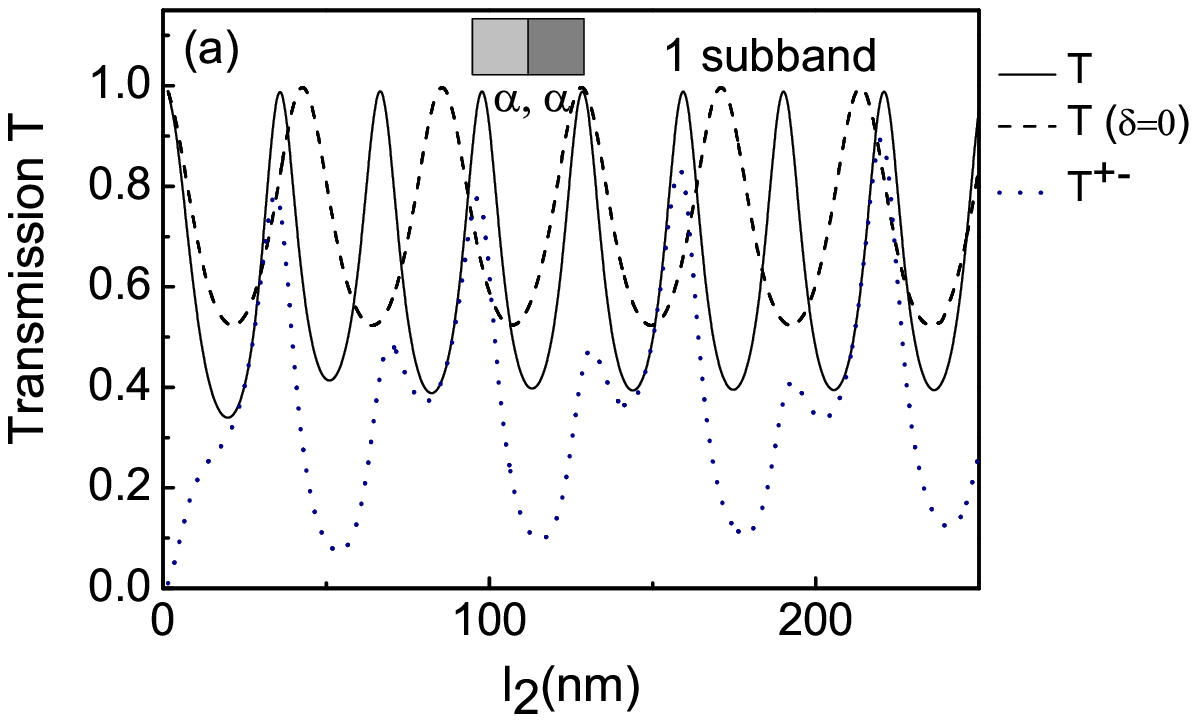}
   \includegraphics[height=6.4cm, width=8cm]{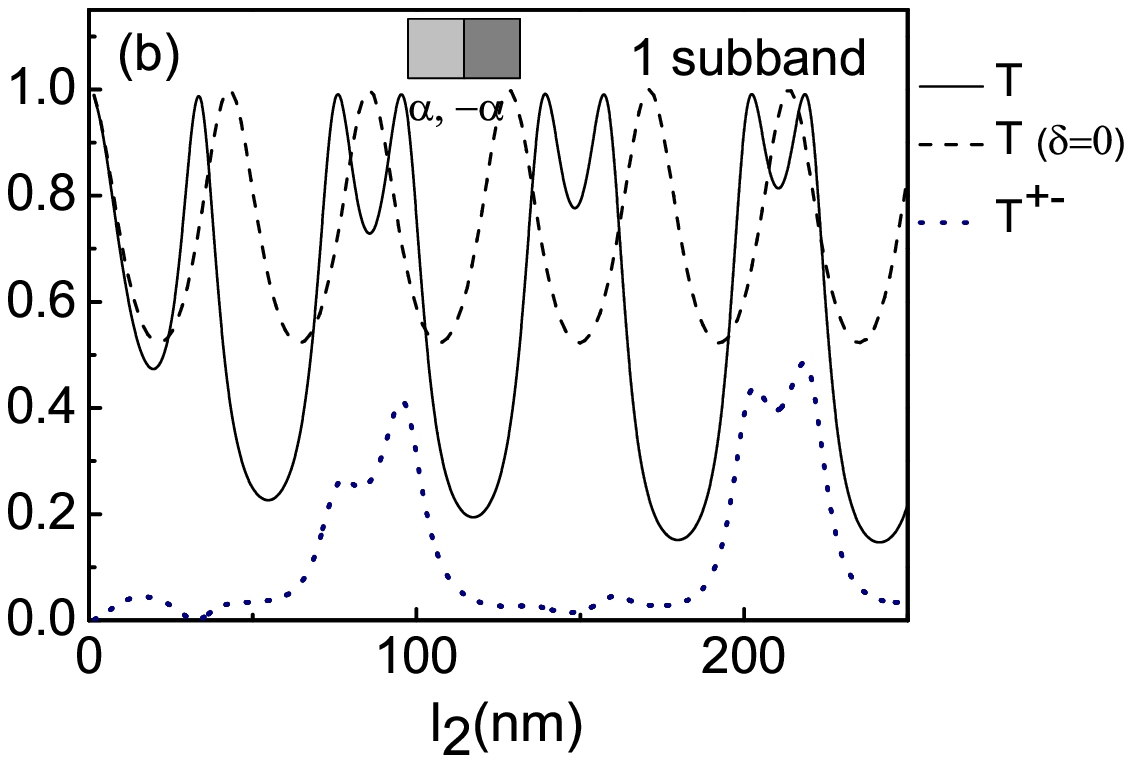}
\end{center}
\vspace*{-1.19cm}
\caption{\label{fig6} (a) Transmission through 
two {\it successive} WG segments, of length $\ell_2$ and strengths $\alpha=\beta=2\alpha_0$, 
versus $\ell_2$. 
The solid curve is the total transmission and the dotted curve the $T^{+-}$ one. 
The dashed curve shows the total transmission for zero mixing. 
  (b) As in (a) but with  
$\alpha=-2\alpha_0$ in the second segment. The energy is $E=E_1+0.2$ meV. }
\end{figure}

In Fig.~\ref{fig4} we plot the total transmission $T$ through $N=3$ units, as a function the length $\ell_2$, for fixed $\ell_1=95$nm and $\alpha_1=\beta_1=0$
in the SOI-free region, see Fig.~\ref{fig1}, 
in two arrangements:  with subband mixing present  $\delta\neq0$  in panel (a) and absent ($\delta=0$) in panel (b). 
For
illustrative purposes here we take 
$\alpha=3\alpha_0$. 
Comparing the two graphs, one notices that the subband mixing introduces aperiodic features mainly in the $T^{+-}$ transmission. Inspecting Fig.~\ref{fig4}(b), one  sees the effect of increasing $\beta$: the minima become deeper and shifted to the  left.

It would be useful to investigate transport through a WG with significantly more than several units. The transmission through $20$ units is shown in Fig.~\ref{fig5}(a) and Fig.~\ref{fig5}(c), as a function of  the ratio $\alpha/\beta$.
 In panels  (a) and (b) the SOI strengths are given in units of $\alpha_0$ and those in panels  (c) and (d) in units of $3\alpha_0$. 
For a clearer comparison the unit lengths $\ell_1=\ell_2=95$ nm and the energy $E=0.13$ meV are the same in all panels. Also, in all cases the SOI is absent in the first segment of the unit.  Only ballistic transport was considered and the aim of 
taking a large number of units, $N$, is to demonstrate the near binary behaviour of the transmission \cite{last}. This is readily seen by contrasting the upper with the lower panels. Notice though that the lower panels involve rather big SOI strengths. 

Apart from the transmission $T$, the conductance $G$ provides valuable information about the nanostructure especially at finite temperatures. $G$ is given by the standard expression 
\begin{equation}
G=\frac{e^2}{h}\int T(E)(-df/dE) dE\,,
\end{equation}
where $f$ is the Fermi-Dirac distribution. In Fig.~(\ref{fig5})(b)

we show $G$ for the same WG, at finite temperature $T_0=0.2$K, as a function of  $\alpha$ 
with $\beta$ kept constant  and equal, respectively, to $\alpha_0$  
 and  $3\alpha_0$ (solid red curves),  and $\beta=0$ (dashed blue curves). 
As can be seen, the dashed and dotted curves coincide since the SOI strength is the same in units of $\alpha_0$.  
The conductance when $\alpha=0$ or $\beta=0$ starts from $2$ since then the WG 
 is completely transparent (no SOI), whereas when both SOI terms 
 are present $G$ starts from a value between $1$ and $2$ due to phase $\phi\neq 0$ and a non-trivial energy dispersion $E(k_y)$, see Eqs.~\ref{energydisp} and (\ref{eigvec2}).  In all cases the first segment of the  unit has zero SOI. For higher values of the SOI strengths, $\beta=3\alpha_0$ (dashed curve in Fig.~\ref{fig5}(d)) the conductance exhibits a more binary behavior. For higher temperatures the qualitative behavior of $G$ remains the same but the dips get shallower. We now assess  the dependence of  the transmission $T$ on the phase $\phi$.
In Fig.~\ref{fig6}  we show $T$ through two successive WG segments, with the same length $\ell_2$ and strength $\beta=2\alpha_0$, versus $\ell_2$. 
 Panel (a) is for $\alpha=2\alpha_0$ and panel (b) for $\alpha=-2\alpha_0$. 
The solid curves show  the total transmission while the dotted ones correspond to $T^{+-}$. For comparison, the values of the total transmission 
for zero mixing ($\delta=0$) are shown as dashed curves. Comparing the two graphs one sees a strong effect the change of the sign of $\alpha$ has on the spin-down 
contribution  which is almost filtered out for 
$\alpha=-2\alpha_0$ and $\ell_2$ smaller than 
about $ 60 $ nm as well as for $\ell_2$ approximately in the range $110$ {\rm nm}$-180$ {\rm nm} .
 
 \begin{figure}[h]
\begin{center}
\setlength{\unitlength}{1cm}%
\hspace*{-0.4cm}  \includegraphics[height=6.4cm, width=6.6cm]{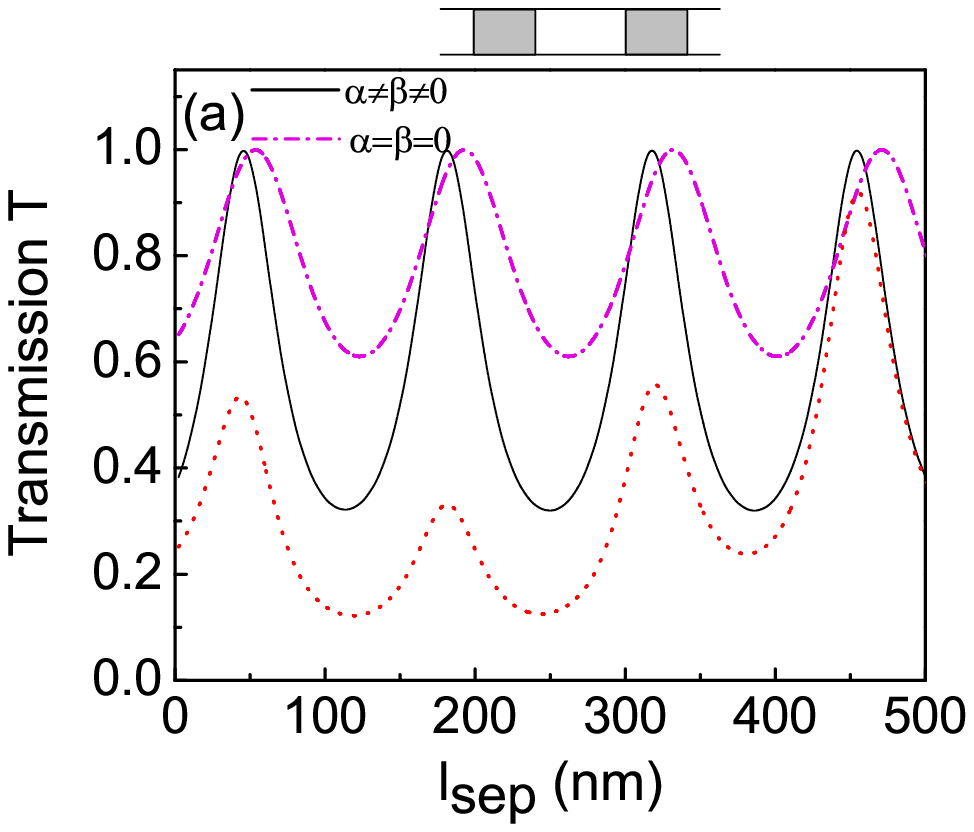}
\hspace*{-0.5cm}
  \includegraphics[height=6.4cm, width=6.6cm]{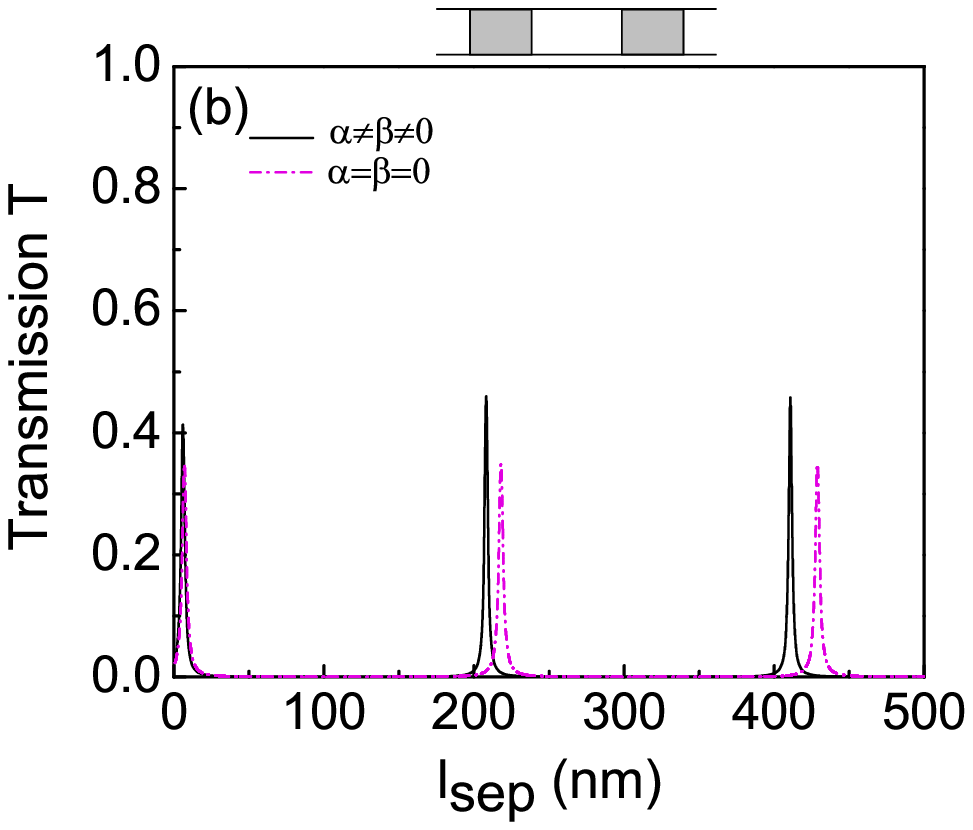}
  \end{center}
\vspace*{-1.2cm}
\caption{\label{fig7} Transmission through two
In$_{x_2}$Ga$_{1-x_2}$As segments, with $\alpha_2=\beta_2=\alpha_0$ and $x_2=0.2$, versus the length $\ell_{sep}$ of a In$_{x_1}$Ga$_{1-x_1}$As  
segment that separates them with $\alpha=0$,  
lengths
$\ell_1=\ell_2=95$ nm, $x_1<x_2$, and $\beta_1=\alpha_0/2$. Panel (a) is for pure GaAs ($x_1=0$)   and a band offset $\Delta V=0.23 {\rm eV}$ and   panel (b) for $x_1=0.1$  and an offset $\Delta V\approx 0.12 {\rm eV}$.   The solid (dash-dotted) curves
show  the total transmission when the SOI is present (absent,  $\alpha_i=\beta_i=0$). In addition, 
 the red dotted curve in   panel (a) shows the $T^{+-}$  transmission.} 
\end{figure}

At this point one may wonder how realistic the difference in SOI strengths is from one region to another or how it can be changed. Firstly, one can use the same material throughout the WG and instead apply gates that can change $\alpha$, from region 1 to region 2, by a factor of 2 to 5. Secondly, if one uses different materials for regions 1 and 2, a band offset exists between them,  
i.e., $V(y)$ is not everywhere zero. As is well known,
the Rashba term is controlled by an external gate and is taken to be zero within the layer made of In$_{x_1}$Ga$_{1-x_1}$As. That is, one can take $\alpha_1\approx 0$ but keep $\beta_1\approx \alpha_0/2$ in the first segment, since it was assumed that $x_1<x_2$. As usual we take $|\alpha_2|=|\beta_2|=\alpha_0$ for 
In$_{x_2}$Ga$_{1-x_2}$As. 
In Fig.~\ref{fig7}(a) we show the case when the two In$_{x_2}$Ga$_{1-x_2}$As segments are separated by a pure GaAs segment ($x_1=0$), free of the Rashba SOI, as a function of the separation length $\ell_{sep}$. For an indium content $x=0.2$, the conduction band mismatch between pure GaAs and In$_x$Ga$_{1-x}$As is experimentally determined \cite{Lu} to be $0.23 {\rm eV}$. In 

Fig.~\ref{fig7}(b) we show the transmission through two 
In$_{x_2}$Ga$_{1-x_2}$As ($x_2=0.2$) segments separated by a RSOI-free segment but now made of In$_{x_1}$Ga$_{1-x_1}$As  with $x_1=0.1$. 

In the case we consider, $x_1< x_2$,  we assume $\Delta V\approx0.12 {\rm eV}$, 
for $x_2=0.2$. The effective masses 
 of In$_{0.2}$Ga$_{0.8}$As and In$_{0.1}$Ga$_{0.9}$As are $0.06$ and $0.064$, respectively    Ê\cite{effmassInGaAs}.
We have taken into account this effective-mass difference between the two materials through the matching of the flux at the interfaces. The solid and dashed-dotted curves show the total transmission with and without SOI present, respectively. As shown, with a band offset present the transmission is smaller. The higher content of indium 
results in narrower transmission peaks when plotted vs $\ell_{sep}$. In both cases, the peak-to-valley ratio is enhanced by the presence of SOI, thus improving 
the performance of a WG as a possible transistor.
We further notice that, although the band offset is large compared to the change in the conduction-band structure caused by the SOI, in the order of a few ${\rm meV}$, the influence of the SOI, through the phase factor in Eq.~(\ref{eigvec2}),  is still important and changes the period of the transmission.

 \begin{figure}[h]
\begin{center}
\vspace{-0.29cm}
\includegraphics[height=5cm, width=5.8cm]{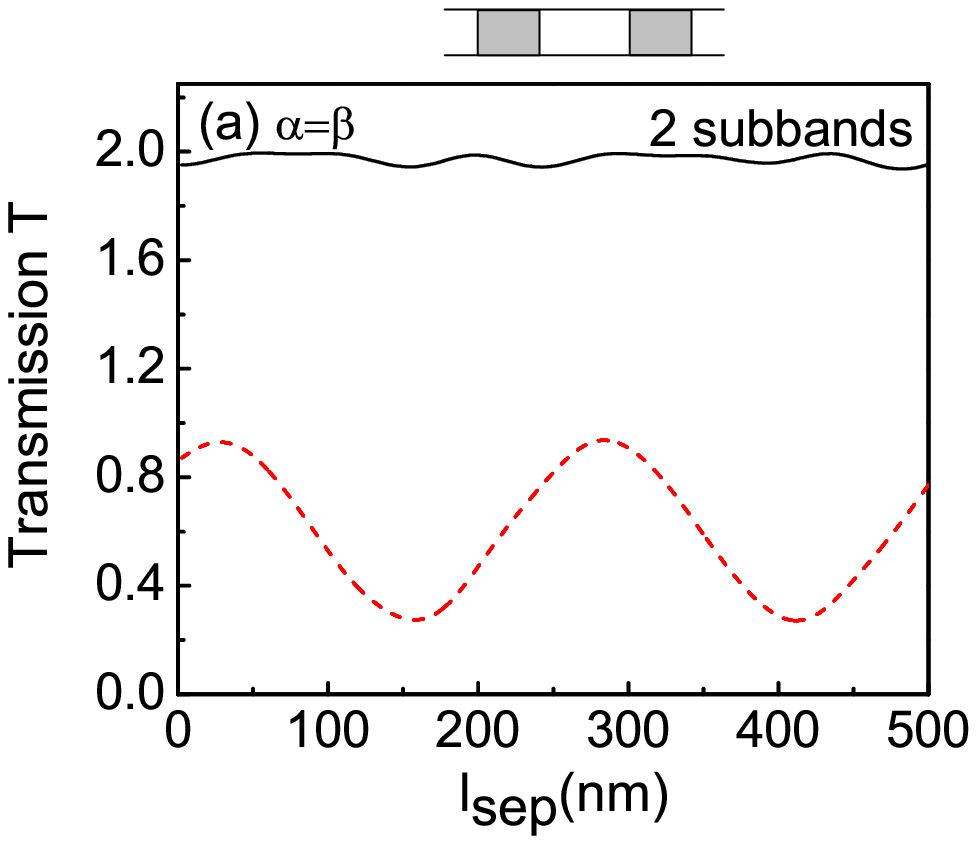}
   \includegraphics[height=5cm, width=5.8cm]{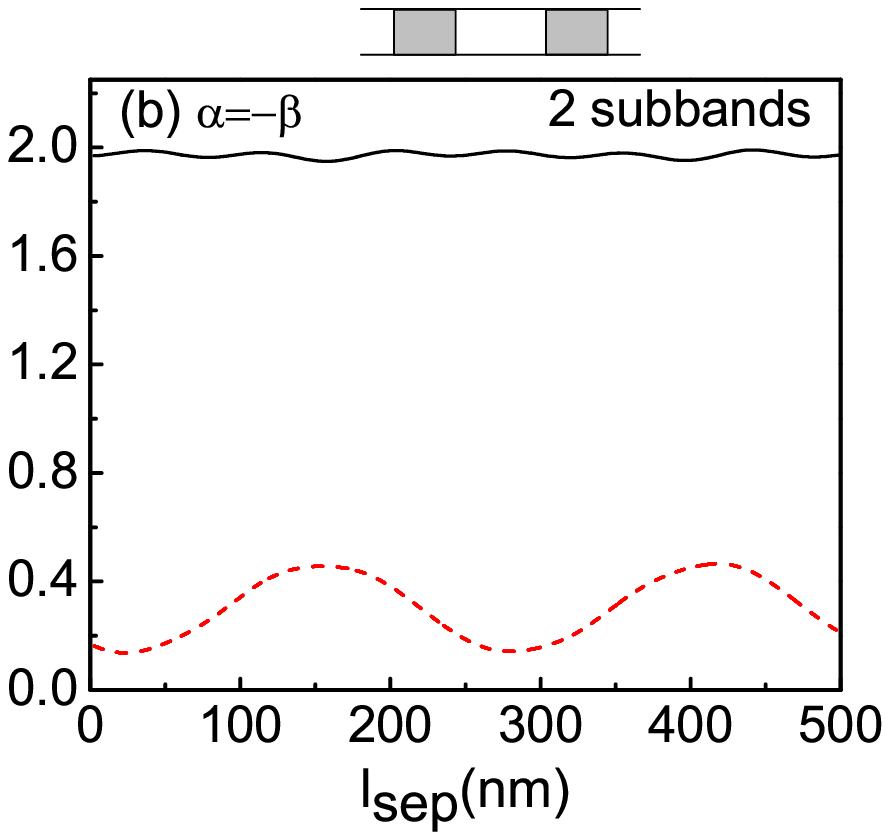}
\end{center}
\vspace*{-0.99cm}
\caption{\label{fig8} (a) Transmission through two {\it simple} units versus their separation $\ell_{sep}$ for fixed $\ell_1=\ell_2=100$ nm  and $\alpha=\beta=\alpha_0$, when 
{\it both subbands} are occupied. The solid curve shows  the total transmission and 
the dashed one the component $T^{+-}$. (b) Same as in (a) but with the second unit having a negative $\alpha$,  i.e., $\alpha=-\alpha_0$.
 The energy is $E=E_1 +1.6$ meV. }
\end{figure}

Next, we consider  the situation when both subbands are occupied, which occurs for  
$E>E_2^{\prime}$. The electron transmission was evaluated through two simple units ($|\alpha|=\beta=\alpha_0$) separated by a SOI-free region of length $\ell_{sep}$. The total transmission $T$ (solid curve) and the transmission from the spin-up to the spin-down state $T^{+-}$ (dashed curve) of the second subband 
are shown in Fig.~\ref{fig8}(a),  when both units have  positive  
strength $\alpha$, and  in Fig.~\ref{fig8}(b) when the second unit has a negative $\alpha$.   
In both cases the total transmission is close to unity, as a result of the high value of the Fermi energy, $E_F=1.6$ meV, while the value of $T^{+-}$ is suppressed when the Rashba coupling changes sign as in the case of only one subband occupied, see Fig.~\ref{fig3}. One can see the filtering effect  only in particular components of the transmission, while the total transmission varies very little with $\ell_{sep}$. We emphasize that the shape and values of $T^{+-}$ are more sensitive to changes in the energy and the strength $\alpha$ than in the case when only one subband is occupied. 

\begin{figure}[h]
\begin{center}
\hspace*{-0.5cm}
   \includegraphics[height=6cm, width=10cm]{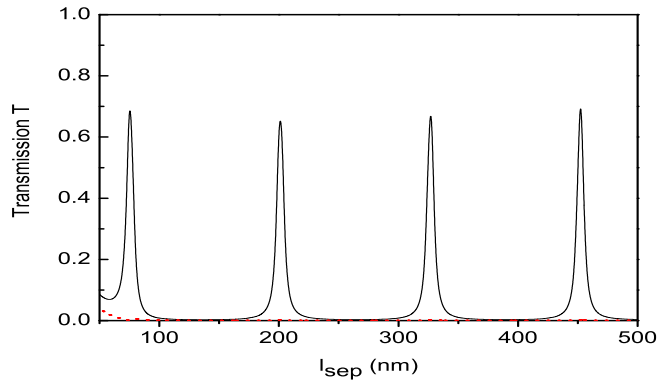}
\end{center}
\vspace*{-1.09cm}
\caption{\label{fig3g110} Transmission through two  
InGaAs segments grown along the $[110]$ direction with $\alpha=\beta=\alpha_0$, versus the length $\ell_{sep}$ of the Rashba SOI-free region that separates them, 
for fixed $\ell_1=\ell_2=95$ nm and  energy  $E=E_1+0.13$ meV.   
The solid curve shows  the total transmission, and 
the  dotted one the $T^{+-}$ transmission.}
\end{figure}

Finally, we consider a WG grown along the $[110]$ direction that was treated theoretically in Sec.~\ref{SecthB}. The Dresselhaus term acquires the simple form
 $-2\beta\sigma_zk_x$ and, as was shown in Sec. III,  
this affects the dispersion relation significantly. We show  
numerical results for the transmission through two InGaAs segments 
separated by a Rashba SOI-free region   
in Fig.~\ref{fig3g110} when only one subband is occupied. The solid curve shows the total transmission and the dotted one, barely visible,   the  $T^{\pm}$ transmission.  . 
As seen, the peak values are noticeably lower than in the previous case and the number of peaks is increased due to the different dispersion relation $E(k_y)$.

\section{Concluding remarks}\label{conc}
We presented results for the electron transmission $T$ through WGs
in which both terms of the SOI are present and  the mixing of the  first two subbands is taken into account. 
In general, the influence of subband mixing is  to shift the (longitudinal) resonances and suppress the spin-down to spin-up transmission. Two growth directions were considered 
$[010]$ and $[110]$, with more attention given to the former.
Further,  changing the sign of the RSOI strength has a very strong filtering effect on the spin-down contribution while leaving the total transmission intact. 

For two segments separated by a SOI-free region, of length $\ell_{sep}$, $T$ shows resonances, as a function of  $\ell_{sep}$, that are most pronounced for $\alpha=\beta$. 
When both subbands are occupied, the total transmission varies little and remains close to unity, so that the filtering effect is contained only in $T^{+-}$. 
Similar to the case when one subband is occupied, changing the sign of the Rashba strength reduces the spin-up to spin-down transmission,  cf. Fig.~\ref{fig8}. 
In addition, we took into account possible band offsets between these segments and the SOI-free region that separates them. As shown in Fig.~\ref{fig7}, this reduces the amplitude of the transmission but does not affect its qualitative dependence on $\ell_{sep}$, 
notice in particular the highly binary structure of the transmission in Fig. 8(b) and consequently that 
of the conductance (not shown) at least for very low temperatures as reflected by Eq. (23).
 
The transmission $T$ and conductance $G$ oscillate as a function of  $\alpha$, $\beta$, or $\alpha/\beta$  if $\alpha$ and  $\beta$ are sufficiently strong. In such a case a nearly square-wave form is shown in Fig.~\ref{fig5}(c) for $T$ and
 in Fig.~\ref{fig5}(d) for $G$ at temperature $T_0=0.2$ K. Both results are in line with those \cite{wan2} for $\beta=0$. For higher temperatures the qualitative behaviour of $G$ remains the same but its maxima are a bit rounded off.
Together with the control of $\alpha$ by a bias \cite{nit} and the independent one of $\beta$ reported very recently \cite{jak}, 
the results indicate that a realistic spin transistor is possible if the SOI-free regions are relatively narrow. 
 
\acknowledgments{}
Our work was supported by the Canadian NSERC Grant No.
OGP0121756.


\begin{references} 
\bibitem{bych}  Y. A. Bychkov and E. I. Rashba, J. Phys. C \textbf{17}, 6039 (1984).

\bibitem{wink} R. Winkler, {\it Spin-orbit coupling effects in two-dimensional electron and hole
systems}, Springer Tracts in Modern Phys. Vol. \textbf{191}  (Springer, New York, 2003).

\bibitem{dres} G. Dresselhaus, Phys. Rev. \textbf{100}, 580 (1955).

\bibitem{nit} J. Nitta, T. Akazaki, H. Takayanagai, and T. Enoki,  Phys. Rev. Lett. \textbf{78}, 1335 (1997); C-M Hu, J. Nitta, T. Akazaki, H. Takayanagi, J. Osaka, P. Pfeffer and W. Zawadzki, Phys. Rev. B \textbf{60}, 7736 (1999).

\bibitem{gate2} G. Engels, J. Lange, T. Sch\"{a}pers and H. L\"uth Phys. Rev. B \textbf{55} R1958 (1997).

\bibitem{gate3} D. Grundler, Phys. Rev. Lett. \textbf{84}, 6074 (2000).

\bibitem{Ivch} E. L. Ivchenko, A. Y. Kaminski, and U. R\"ossler, Phys. Rev. B \textbf{54}, 5852 (1996).

\bibitem{Krebs} O. Krebs, D. Rondi, J. L. Gentner, L. Goldstein, and P. Voisin, Phys. Rev. Lett. \textbf{80}, 5770 (1998).

\bibitem{Gan2} S. D. Ganichev, V. V. Bel'kov, L. E. Golub, E. L. Ivchenko, P. Schneider, S. Giglberger, J. Eroms, J. De Boeck, G. Borghs, W. Wegscheider, D. Weiss, and W. Prettl, 
Phys. Rev. Lett. \textbf{92}, 256601 (2004); V. I. PerelÕ, S. A. Tarasenko,  I. N. Yassievich,
S. D. Ganichev, V. V. BelÕkov, and W. Prettl,  Phys. Rev. B  \textbf{67}, 201304(R) (2003).

\bibitem{Datta} S. Datta and B. Das, Appl. Phys. Lett. \textbf{56}, 665 (1990).

\bibitem{schl1} J. Schliemann, J. C. Egues, and D. Loss, Phys. Rev. Lett. \textbf{90}, 146801 (2003).

\bibitem{SFET_DSO} S. Bandyopadhyay and M. Cahay, Appl. Phys. Lett. \textbf{85}, 1814 (2004).

\bibitem{zut} I. \v{Z}uti\'c, J. Fabian, and S. Das Sarma, Rev. Mod. Phys. \textbf{76}, 323 (2004).

\bibitem{wan1} X. F. Wang, P. Vasilopoulos, and F. M. Peeters, Phys. Rev. B \textbf{65}, 165217 (2002).

\bibitem{wan2} X. F. Wang and P. Vasilopoulos, Appl. Phys. Lett. \textbf{83}, 940 (2003).

\bibitem{jak} J. D. Koralek, C. Weber, J. Orenstein, A. Bernevig, S. Zhang, S. Mack, and D. Awschalom, Nature \textbf{458}, 610 (2009).

\bibitem{Golub} N. S. Averkiev and L. E. Golub, Phys. Rev. B \textbf{60}, 15582 (1999).

\bibitem{Wangsubbmix} X. F. Wang and P. Vasilopoulos, Phys. Rev. B \textbf{68}, 035305 (2003).

\bibitem{Dres110} X. Cartoix\`a, L.-W. Wang, D. Z.-Y. Ting, Y.-C. Chang, Phys. Rev. B \textbf{73}, 205341 (2006); B. A. Bernevig, J. Orenstein, and S.-C. Zhang, Phys. Rev. Lett. \textbf{97}, 236601 (2006);
M.-H. Liu, K.-W. Chen, S.-H. Chen, and C.-R. Chang, Phys. Rev. B \textbf{74}, 235322 (2006).

\bibitem{anothWang} M. Wang, K. Chang, and K. S. Chan, Appl. Phys. Lett. \textbf{94}, 052108 (2009).

\bibitem{mol} L. W. Molenkamp, G. Schmidt, and G. E. W. Bauer, Phys. Rev. B  \textbf{64}, 121202(R) (2001).

\bibitem{Xu} H. Xu, Phys. Rev. B  \textbf{47}, 9537 (1993).

\bibitem{last}  For $N=20$ the total length of the structure is $L_{tot}= 2\mu m$ and comparable to the phase relaxation length $\ell_{\phi}$ which is about $0.7\mu m$, see Choi {\it et al.} Phys. Rev. B {\bf 36}, 7751 (1987). For larger $N$  the implicit assumption $L_{tot}\leq \ell_{\phi}$ will not hold
but one could reduce the unit length to satisfy this condition.

\bibitem{Lu} L. Lu, J. Wang, Y. Wang, W. Ge, G. Yang, and Z. Wang, J. Appl. Phys. \textbf{83}, 2093 (1998).

\bibitem{effmassInGaAs} J. C. Fan and Y. F. Chen, J. Appl. Phys. \textbf{80}, 6761 (1996).

\end{references}
\end{document}